\documentclass[letterpaper, 10 pt, conference]{ieeeconf}  

\IEEEoverridecommandlockouts                              

\usepackage{graphics}
\usepackage{multirow}
\usepackage{graphicx}
\usepackage{float}
\usepackage{siunitx}
\usepackage{array,booktabs,arydshln,xcolor}
\usepackage[caption=false]{subfig}

\title{\LARGE \bf
Attention Monitoring and Hazard Assessment with Bio-Sensing and Vision: Empirical Analysis Utilizing CNNs on the KITTI Dataset
}

\author{Siddharth and Mohan M. Trivedi
\thanks{Siddharth and Mohan M. Trivedi are with The Laboratory for Intelligent and Safe Automobiles (LISA) at the University of California San Diego, La Jolla, CA, USA, 92093.
        {\tt\small ssiddhar@eng.ucsd.edu, mtrivedi@eng.ucsd.edu}}%
}

\begin{document}

\maketitle
\thispagestyle{empty}
\pagestyle{empty}

\begin{abstract}
Assessing the driver's attention and detecting various hazardous and non-hazardous events during a drive are critical for driver's safety. Attention monitoring in driving scenarios has mostly been carried out using vision (camera-based) modality by tracking the driver's gaze and facial expressions. It is only recently that bio-sensing modalities such as Electroencephalogram (EEG) are being explored. But, there is another open problem which has not been explored sufficiently yet in this paradigm. This is the detection of specific events, hazardous and non-hazardous, during driving that affects the driver's mental and physiological states. The other challenge in evaluating multi-modal sensory applications is the absence of very large scale EEG data because of the various limitations of using EEG in the real world. In this paper, we use both of the above sensor modalities and compare them against the two tasks of assessing the driver's attention and detecting hazardous vs. non-hazardous driving events. We collect user data on twelve subjects and show how in the absence of very large-scale datasets, we can still use pre-trained deep learning convolution networks to extract meaningful features from both of the above modalities. We used the publicly available KITTI dataset for evaluating our platform and to compare it with previous studies. Finally, we show that the results presented in this paper surpass the previous benchmark set up in the above driver awareness-related applications.
\end{abstract}

\section{Introduction}
With the development of increasingly intelligent and autonomous vehicles it has been possible to assess the criticality of a situation much before the event actually happens. It has also become possible to monitor the driver's responses to various events during the drive. While computer vision continues to be the preferred sensing modality for achieving the goal of assessing driver awareness, the use of bio-sensing systems in this context has received wide attention in recent times \cite{DrAw_1,DrAw_2}. Most of these studies have used electroencephalogram (EEG) as the preferred bio-sensing modality. While these studies have shown that EEG can prove to be very useful for assessing fatigue and attention in the driving context, it generally suffers from low spatial resolution. Furthermore, the use of high-density EEG systems is impractical in a real-world driving context.

Driver awareness depends highly on the driver's physiology since different people react differently to fatigue and to their surroundings. This means that a single fit-for-all type of approach using computer vision based on eye blinks/closure etc. might not scale very well across drivers. It is here that the use of EEG may play a useful role in assessing driver awareness by continuously monitoring the human physiology. Furthermore, EEG may prove to be very useful for detecting hazardous vs. non-hazardous situations on short time scales (such as 1-2 seconds) if such situations do not register in the driver's facial expressions in such short time periods.

The advent of deep learning has translated very well towards vision-based systems for many applications, among them, driver behavior and attention monitoring \cite{DL_Driver1,DL_Driver2}. But, these advances have not translated towards the data from bio-sensing modalities such as EEG. This is primarily due to the difficulty in collecting very scale bio-sensing data. Collecting bio-sensing data on a large scale is costly, laborious, and time-consuming,  whereas for collecting image/videos even a smartphone's camera may suffice. Hence, we explore the use of pre-trained image-based deep learning networks to extract meaningful features for both sensor modalities.

This study focuses on driver awareness and his/her perception of hazardous/non-hazardous situations from bio-sensing as well as vision-based perspectives. We utilize the KITTI dataset \cite{KITTI} for evaluation and individually use features from EEG and image data to compare the performance of these modalities. We also use the fusion of features and show how in certain cases, the use of multiple modalities may be advantageous. We also show how pre-trained deep neural networks can be utilized to extract features from these modalities for boosting the performance even in the absence of very large scale data. To the best of our knowledge, this study is the most comprehensive view of using EEG and vision modalities towards assessing driver awareness and hazard assessment. Finally, we would like to emphasize that the data collection set up used in this study is very practical to use in ``real-world'' i.e. it is compact in design, wireless, and comfortable to use for prolonged time intervals.

\section{Related Studies}
Driver monitoring for assessing attention, awareness, behavior prediction, etc. has usually been done using vision as the preferred modality \cite{DriverAttention_Review,LISA_1,LISA_2}. This is carried out by monitoring the subject's facial expressions and eye-gaze \cite{LISA_4,LISA_5} which are used to train machine learning models. But, almost all such studies utilizing ``real-world'' driving scenarios have been conducted during daylight when ample ambient light is present. Even if infra-red cameras are used to conduct such experiments at night, vision modality suffers from occlusion and widely varying changes in illumination \cite{DriverAttention_Review}, both of which are not uncommon in driving scenarios. 

Hazard assessment is a problem that has been tackled with vision as the primary modality \cite{Kitti_Dataset_Study} and evaluated on the KITTI dataset \cite{KITTI}. Additionally, it has also been shown that the use of EEG can classify hazardous vs. non-hazardous situations over short time periods which is not possible with images/videos \cite{Freiburg_EEG}. We take this recent study on the KITTI dataset as the benchmark for our evaluation. The KITTI dataset contains vision data (along with other sensors) collected in rural areas and on highways in Karlsruhe city of Germany. The dataset contains many driving scenarios with up to 15 cars and 30 pedestrians visible per image. This makes it ideal for our study since the scene complexity in the dataset varies a lot more than many other such datasets. 

The study \cite{Freiburg_EEG} suffers from three major limitations. First, the study superimposes non-existent stimulus over the real-world driving images which introduce an uncanny aspect since such stimuli are not what one expects during a real-world drive. Second, the study does not use the most common sensor modality to assess the driver's awareness i.e. vision. Hence, no baseline comparison can be done between the two kinds of sensor modalities. Third, the EEG features used in the research are the most commonly used ones and are in no way tuned to the specific application at hand. No attempt has been made to extract higher-level EEG features that are more relevant to human cognition. Finally, the study contains a pool of only five subjects which is quite small since the usual norm is to use at least ten subjects for bio-sensing based applications.

\section{Research Methods}
In this section, we discuss the various research methods that we employed to pre-process the data and extract features from each sensor modality used in this study.

\subsection{EEG-based Feature Extraction}
The cognitive processes pertaining to attention and mental load such as while driving are not associated with only one part of the brain. Hence, our goal was to map the interaction between various regions of the brain to extract relevant features related to attention. The EEG was initially recorded from 14-channel Emotiv EEG headset at 128 Hz sampling rate. The locations of the EEG channels according to the International 10-20 system were: AF3, AF4, F3, F4, F7, F8, FC5, FC6, P7, P8, T7, T8, O1, and O2. This EEG headset was chosen because it is compact, wireless, and easy to use in real-world settings. We first pre-processed the data using EEGLAB \cite{EEGLAB} toolbox. We used the artifact subspace reconstruction (ASR) pipeline in the toolbox to remove artifacts related to eye blinks, movements, line noise, etc. \cite{EEG_ASR}. Then, the cleaned EEG data was band-pass filtered between 4-45 Hz to preserve the data from the most-commonly used EEG frequency bands. We then employed two distinct and novel methods to extract EEG features that captured the interplay between various brain regions to map human cognition.

\subsubsection{Features based on Mutual Information}
To construct the feature space that can map the interaction of EEG information between various regions of the brain, we calculated the mutual information between signals from different parts of the brain \cite{conditional_entropy, my_TBME}. Such features were opted for since they measure the changes in EEG across the various regions of the brain which might be more expressive of human cognition rather than spatially local features. The mutual information $I(X;Y)$ of discrete random variables $X$ and $Y$ is defined as 

\begin{equation}
I(X;Y) = \sum_{x\in X}\sum_{y\in Y}p(x,y) log\bigg(\frac{p(x,y)}{p(x)p(y)}\bigg)
\end{equation}

The desired feature of conditional entropy $H(Y|X)$ is related to the mutual information $I(X;Y)$ by

\begin{equation}
I(X;Y) = H(Y) - H(Y|X)
\end{equation}

We calculated the conditional entropy using mutual information between all possible pairs of EEG electrodes for a given trial. Hence, for 14 EEG electrodes, we calculated 91 EEG features based on this measure.

\subsubsection{Features based on Deep Learning}
The most commonly used EEG features are the power-spectrum density (PSD) of different EEG bands. But, these features in themselves do not take into account the EEG-topography i.e. the location of EEG electrodes. Hence, we try to exploit EEG-topography to extract information regarding the interplay between different brain regions. We also wanted to utilize pre-trained convolution networks since training a deep neural network from scratch needs a very large amount of data which is difficult to acquire. Thus, we sought to convert time-domain EEG data to image domain for utilizing such pre-trained convolution networks. 

We calculated the PSD of three EEG bands namely theta (4-7 Hz), alpha (7-13 Hz) and beta (13-30 Hz) for all the EEG channels. The choice of these three specific EEG bands was made since they are the most-commonly used EEG bands and have been shown in multiple studies to contribute significantly towards human cognition. We averaged the PSD for each band thus calculated over the complete trial. These features from different EEG channels were then used to construct 2-D EEG-PSD heatmap for each of the three EEG bands using bicubic interpolation. These heat-maps now contain the information related to EEG topography in addition to spectrum density at every location \cite{my_EMBC}.

\begin{figure}[!ht]
\begin{tabular}{cc}
\includegraphics[width=\linewidth]{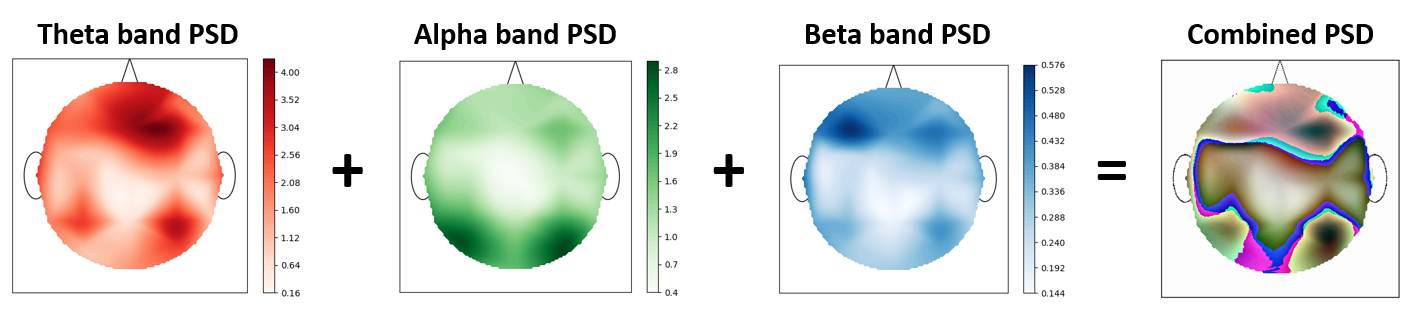} 
\end{tabular}
\caption{PSD heat-maps of the three EEG bands i.e. theta (\textcolor{red}{red}), alpha (\textcolor{green}{green}), and beta (\textcolor{blue}{blue}) EEG bands are added according to respective color-bar range to get combined RGB heat-map image.(Circular outline, nose, ears, and color-bars have been added for visualization only.)}
\label{fig:eeg-bands}
\end{figure}

Fig. \ref{fig:eeg-bands} shows an example of these 2-D heatmaps for each of the three EEG bands. As can be seen from the figure, we plot each of the three EEG bands using a single color channel i.e. red, green and blue. We then add these three color band images to get a color RGB image containing information from the three EEG bands. The three color band images are added in proportion to the amount of EEG power in the three bands using alpha blending \cite{MatPlotLib} by normalizing each band's power using the highest value in the image. Hence, following this procedure we are able to represent the information in the three EEG bands along with topography through a single color image. The interaction between these three colors (thus forming new colors by adding the three primary colors) in various quantities is representative of the information about the distribution of spectral power across the brain.

This combined colored image representing EEG-PSD with topography information is then fed to a pre-trained deep-learning based VGG-16 convolution neural network \cite{VGG} to extract features from this image. This network consists of 16 weight layers and has been trained with more than a million images for 1,000 object categories using the Imagenet Database \cite{ImageNet}. Previous research studies \cite{VGG_2,my_itsc} have shown that using features from such ``off-the-shelf" neural network can be used for various classification problems with good accuracy. The EEG-PSD colored image is resized to $224{\times}224{\times}3$ for input to the network. The last layer of the network classifies the image into one of the 1000 classes but since we are only interested in ``off-the-shelf'' features, we extract 4,096 most significant features from the last but one layer of the network. The EEG features from this method are then combined with those from the mutual information method for further analysis.

\subsection{Facial expression-based feature extraction}
The analysis of facial expressions has been the preferred modality for driver attention analysis. Most of the research work in this area has been done by tracking fixed localized points on the face based on face action units (AUs).

First, we extracted the face region from the frontal body image of the person captured by each camera frame. This was done by fixing a threshold on the image size to reduce its extreme ends and placing a threshold of minimum face size to be $50{\times}50$ pixels. This was done to remove any false positives and decrease the computational space for face detection. We then used the Viola-Jones object detector with Haar-like features \cite{Viola_Jones} to detect the most likely face candidate. In extremely uncommon cases when the face detector failed due to major occlusion by the subject's hands in front of their face, the frames were discarded.

\begin{figure}[!ht] \centering
{\includegraphics[width=3.45in]{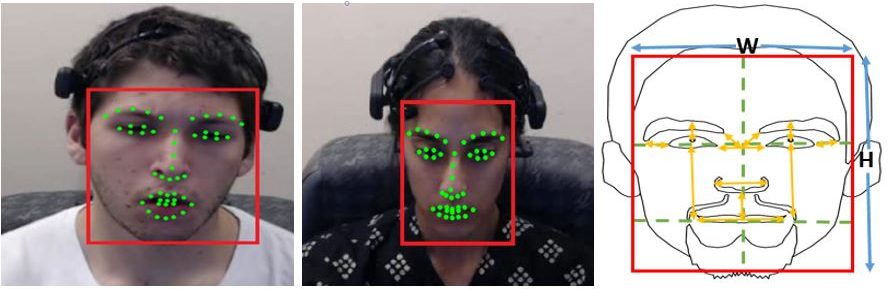}}
\vspace{-15pt}
\caption{Detected face (marked in \textcolor{red}{red}) and face localized points (marked in \textcolor{green}{green}) for two participants (left and center) in the study, and some of the features (marked in \textcolor{yellow}{yellow}) computed using the coordinates of the face localized points.}
\label{fig:face-features}
\end{figure}

\begin{figure*}[!ht] \centering
{\includegraphics[width=\textwidth, height = 1.7in]{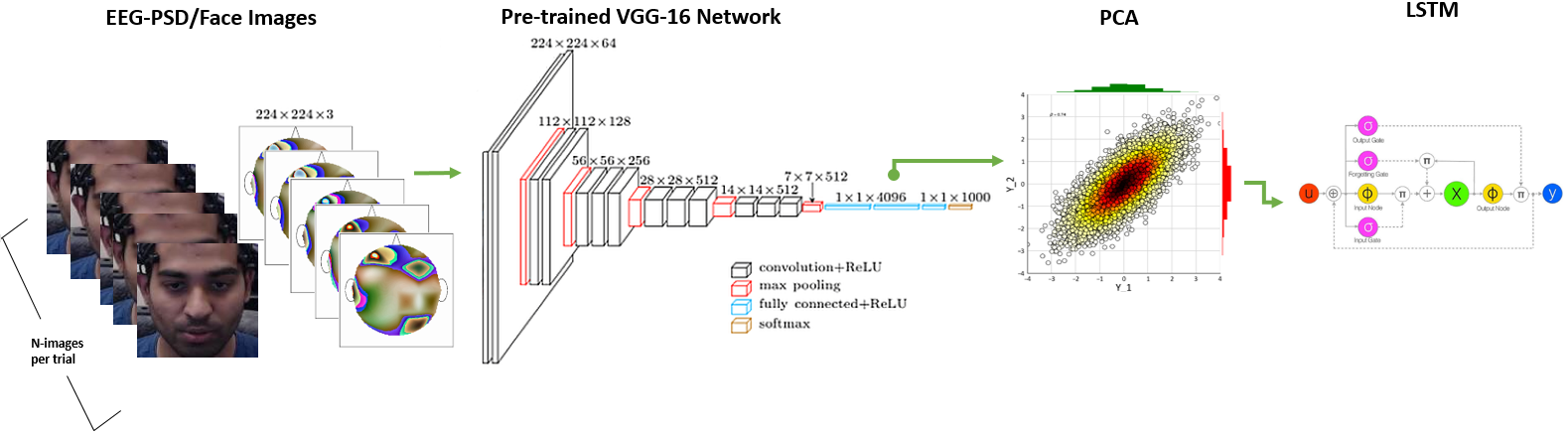}}
\vspace{-15pt}
\caption{Network architecture for EEG-PSD trend based Deep Learning method.}
\label{fig:eeg_method_6}
\end{figure*}

\subsubsection{Facial points localization based features}
Face action units (AUs) has been used for a variety of applications ranging from affective computing to face recognition \cite{face_AUs}. Our goal was to use face localized points similar to AUs without identifying facial expressions such as anger, happiness, etc. since they are not highly relevant in driving domain and short time intervals. We applied the state-of-the-art Chehra algorithm \cite{chehra} to the extracted face candidate region from above. This algorithm outputs the coordinates of 49 localized points representing various features of the face as in Fig. \ref{fig:face-features}. The choice of this algorithm was done because of its ability to detect these localized points through its pre-trained models and hence not needing training for any new set of images. These face localized points are then used to calculate 30 different features based on the distances such as between center of the eyebrow to the midpoint of the eye, between the midpoint of nose and corners of the lower lip, between the midpoints of two eyebrows, etc. and angles between such line segments. To remove variations by factors such as distance from the camera and face tilt, we normalized these features using the dimensions of the face region. To map the variation in these features across a trial, we calculated the mean, $95^{th}$ percentile (more robust than maximum), and standard deviation of these 30 features across the frames in the trial. In this manner, we computed 90 features based on face-localized points from a particular trial.

\subsubsection{Deep Learning-based features}
For the extraction of deep learning-based features, we used the VGG-Faces networks instead of VGG-16 \cite{VGG_Faces}. This was done to extract features more relevant to faces since the VGG-Faces network has been trained on more than 2.6 million face images from more than 2,600 people rather than on various object categories in the VGG-16 network. Similar to the feature extraction methods above, we sent each face region part to the network and extract the most significant 4,096 features. To represent the changes in these features across the trial i.e. across the frames, we calculated the mean, $95^{th}$ percentile, and standard deviation of the features across the frames in a trial. The features from the above two methods were then combined for further analysis.

\subsection{Assessing trends of sensor features using Deep Learning}
The features discussed in sections III.A and III.B above were computed over the whole trial such as by generating a single EEG-PSD image for a particular trial. This is a special case when the data from the whole trial is being averaged. Here, we propose a novel method to compute the trend of these features i.e. their variation in a trial based on deep learning. To compute features with more resolution we generated multiple EEG-PSD images for successive time durations in a trial. Fig. \ref{fig:eeg_method_6} shows the network architecture for this method. The EEG-PSD images are generated for multiple successive time durations in a trial each of which is then sent to the VGG-16 network to obtain 4,096 most significant features. Similarly, this process was done for conditional entropy features by calculating this over multiple time periods in a trial rather than once on the whole trial. We then use principal components analysis (PCA) \cite{PCA} to reduce the feature size to 60 to save computational time in the next step. These $60{\times}N$ ($N =$ number of successive time intervals) features were then sent as input to a Long Short Term Memory (LSTM) network \cite{LSTM}. The same process was performed for face-based features.

The LSTM treats each of these features as a time-series and is trained so as to capture the trend in each of them for further analysis. This method could only be applied when the time duration of the trials is fixed since the length of each time series should be the same. Since the duration of KITTI videos varied widely, we applied this method only in the trials used for detecting hazardous/non-hazardous situations.

\begin{figure}[!ht] \centering
\includegraphics[width=\linewidth, height=1.7in]{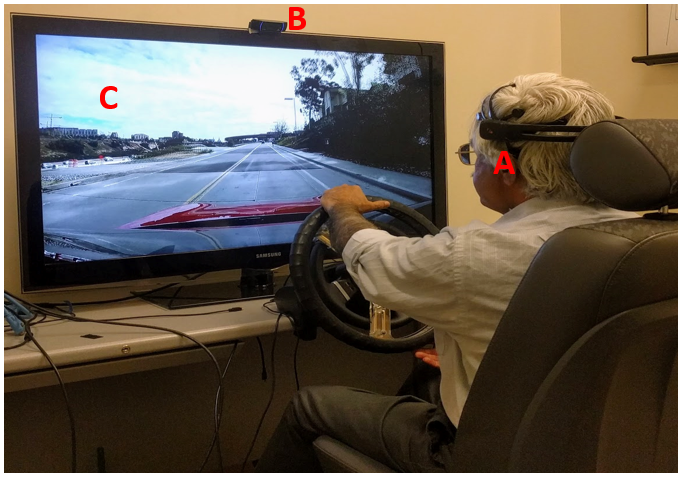} 
\caption{Experiment setup for multi-modal data collection. (A) EEG Headset, (B) External camera, and (C) Driving videos displayed on the screen. The subject sits with her/his arms and feet on a driving simulator with which s/he interacts while watching the driving videos.}
\label{fig:exp_setup}
\end{figure}

\section{Dataset Description}
In Fig. \ref{fig:exp_setup} we show the experimental setup for data collection with driving videos used as the stimulus in our dataset. The participants are comfortably seated equipped with an EEG headset (Emotiv EPOC). Facial expressions of each subject are recorded using a camera in front of him/her. The participants are asked to use a driving simulator which they are instructed to control as per the situation in the driving stimulus. For example, if there is a ``red light" or ``stop sign" at any point in a driving stimulus video, the participants should press and hold the brake. Twelve participants (most of them in their 20s with two older than 30 years) based in San Diego participated in our study. All the modalities were synchronized together using Lab Streaming Layer (LSL) software framework \cite{LSL}.

\begin{figure}[!ht] \centering
{\includegraphics[width=\linewidth, height = 1.7in]{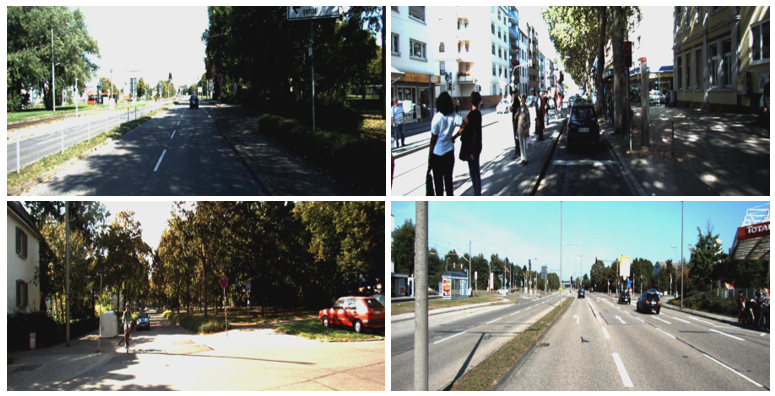}}
\caption{Various image instances with varying illumination conditions and type of road (street, single-lane, highway, etc.) from the KITTI Dataset.} 
\label{fig:dataset_image}
\end{figure}

For consistency between our work and other previous studies \cite{Freiburg_EEG,Kitti_Dataset_Study}, we used 15 video sequences from the KITTI dataset \cite{KITTI}. These video sequences range from 14 to 105 seconds. These videos in the dataset were recorded at $1242{\times}375$ resolution at 10 frames-per-second (fps). We resized the videos to $1920{\times}580$ to fit the display screen in a more naturalistic manner. Two external annotators marked the above 15 video sequences as requiring low-driver attention or high-driver attention based on the context in the video. For example, the video instances where the car is not moving at all were characterized as low attention instances whereas driving through narrow streets with pedestrians on the road were labeled as instances with high driver attention required. These sequences contained videos from widely varying illumination and driving conditions (in street, highway, city, etc.) as shown in Fig. \ref{fig:dataset_image}.

\begin{figure}[!ht] \centering
{\includegraphics[width=3.45in]{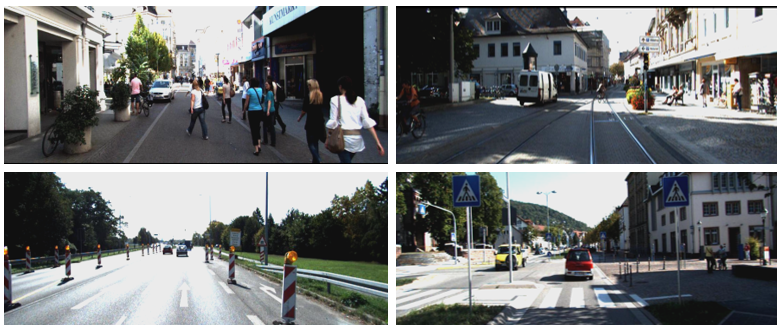}}
\vspace{-15pt}
\caption{(A) Examples of 2-seconds incidents classified as hazardous. Examples include pedestrians crossing the street without a crosswalk while the ego vehicle is being driven and another vehicle overtaking suddenly. (B) Examples of 2-seconds incidents classified as non-hazardous. Examples include stop signs and railway crossing signs.}
\label{fig:incidents-examples}
\end{figure}

Second, 41 instances, each two second long were characterized as hazardous/non-hazardous. Fig. \ref{fig:incidents-examples} presents some examples of instances from both categories. As an example, a pedestrian suddenly crossing the road ``unlawfully'' or a vehicle overtaking suddenly represents hazardous events whereas ``red'' traffic sign at a distance and a pedestrian at a crossing with ego vehicle not in motion are examples of non-hazardous events. Among these instances, 20 instances were labeled as hazardous whereas rest as non-hazardous. Hence, the goal is to classify such instances in a short time period of two seconds using the above sensor modalities.

\section{Quantitative analysis of EEG and vision sensor modalities}
In this section, we present the single modality and multi-modal evaluation results for driver attention analysis and hazardous/non-hazardous instances classification. For each modality, we first used PCA \cite{PCA} to reduce the number of features from the above algorithms to 30. We then used extreme learning machines (ELM) \cite{ELM} for classification. These features were normalized between -1 and 1 across the subjects before training. A single hidden layer ELM was used with triangular basis function for activation. For the method with trend based temporal EEG and face features data, we used two layer LSTM with 200 and 100 neurons in respective layers. The network training was done using stochastic gradient descent with a momentum (SGDM) optimizer.

We performed leave-one-subject-out cross-validation for each case. This meant that the data from 11 subjects was used for training at a time and the classification was done on the remaining $12^{th}$ subject. This choice of cross-validation was driven by two factors. First, this method of cross-validation is much more robust and less prone to bias than models such as leave-one-sample-out cross-validation that constitutes training data from all the subjects at any given time. Second, since the data contained 180 trials only, as opposed to thousands of trials for any decent image-based deep-learning dataset, it did not make sense to randomly divide such a small number of trials to training, validation, and test sets since it might introduce bias by uneven division across trials from individual subjects.

\subsection{Evaluating attention analysis performance}
In this section, we evaluate single and multi-modal performance for assessing the driver's attention across the video trials. For all the four modalities, the features as defined above were calculated for data from each video trial. The ELM-based classifier was then trained based on each video trial divided into one of the two classes representing low-attention and high-attention required by the driver.

\begin{figure}[!ht] \centering
{\includegraphics[width=3.45in, height=1.5in]{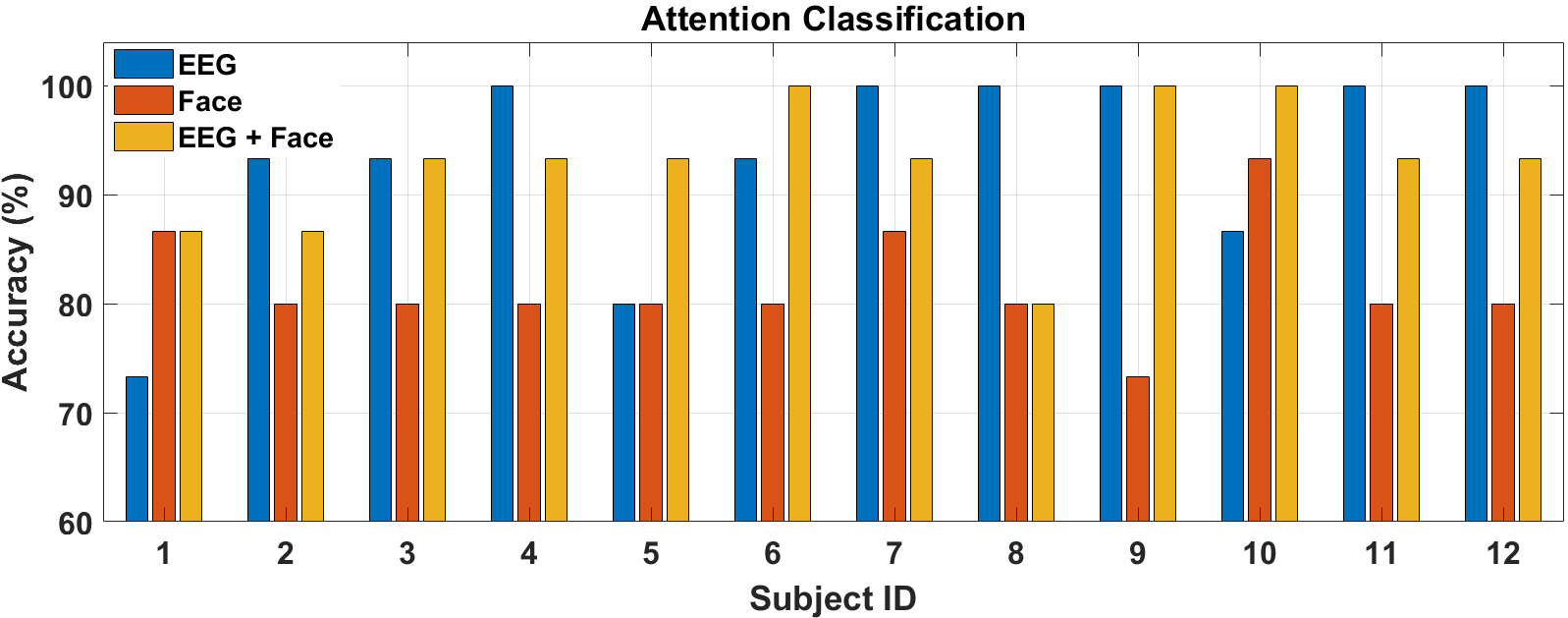}}
\vspace{-15pt}
\caption{Single modality classification performance for driver attention analysis.}
\label{fig:attention-evaluation}
\end{figure}

Fig. \ref{fig:attention-evaluation} shows the result for each modality for attention classification. The mean performance across the subjects for EEG, faces, and EEG + faces combined are 93.33 \%, 81.67\%, and 92.78\% respectively. Clearly, EEG performs better than vision modality, so much so that adding the features from the two together doesn't lead to an increase in the performance. We also see that across most of the subjects, EEG performs better than vision modality. This is perhaps because the slowly-varying facial expressions during a driving task are not so expressive as to map the driver's attention very well. It is also to be noted that for all the cases, the accuracy of our system is significantly higher than chance (50\% accuracy). We also note that the accuracy between subjects varies between EEG and faces modality. This may be due to some subjects being facially more expressive than others in the driving scenario.

\subsection{Evaluating hazardous/non-hazardous incidents classification}
In this section, we present the results of the evaluation of the modalities over very short time intervals (2 seconds) pertaining to hazardous/non-hazardous driving incidents as shown in Fig. \ref{fig:incidents-examples}. Since it is not possible for the subjects to tag the incidents while they are participating in the driving simulator experiment and hence these incidents were marked by the external annotators as mentioned above in Section IV. 

\begin{figure}[!ht] \centering
{\includegraphics[width=3.45in, height=1.5in]{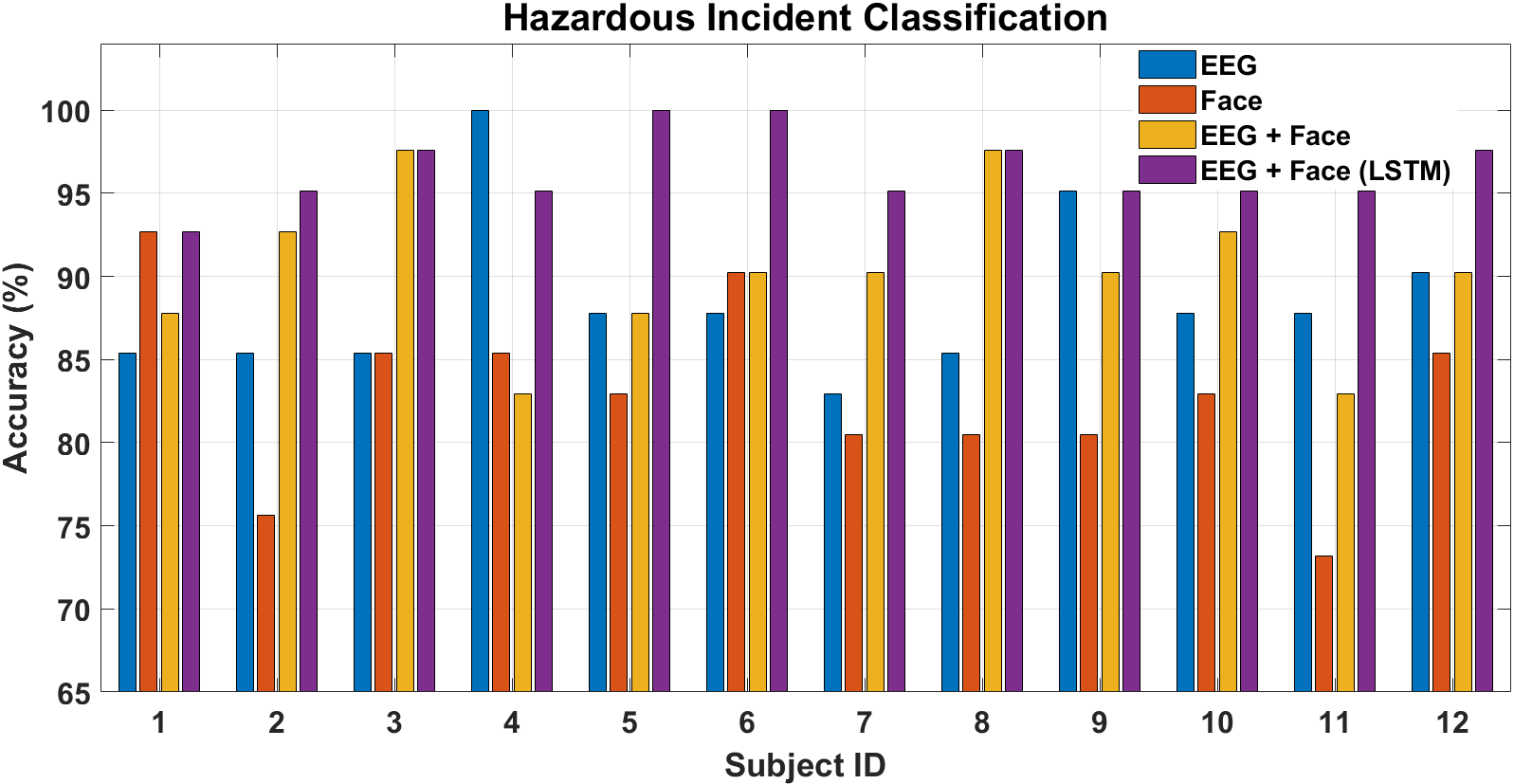}}
\vspace{-15pt}
\caption{Multi-modality classification performance for driver attention analysis.}
\label{fig:hazard-evaluation}
\end{figure}

The mean accuracies across the subjects for EEG, faces, EEG + faces were 88.41\%, 82.93\%, and 90.24\%. This shows that again EEG performs better than vision modality but more importantly, the combined performance of these modalities is better than individual ones. This means that on very short time periods (such as 2 seconds here), face features contain information independent and additive to what EEG features have. It is also to be noted that the performance of EEG in our feature extraction and processing pipeline is significantly better than the previous hazard analysis study (AUC 0.79) \cite{Freiburg_EEG}. Also, again we see that the accuracies for all cases are significantly higher than chance level.

We also note that on performing analysis with finer temporal resolution using LSTMs, the performance further increases to an average of 96.34\%. This shows how the type of traditional and deep learning features we extracted above can be used for such a task with a good performance by utilizing multiple sensor modalities. 

\section{Concluding Remarks}
The use of bio-sensing modalities combined with audio-visual ones is rapidly expanding. With the advent of compact bio-sensing systems capable of data collection during real-world tasks such as driving, it is natural that this research area will gather more interest in the coming years. In this work, we showed how a commercially available compact EEG system with the capability of easily collecting data in the driving scenario can be used to monitor driver's attention in both short and long periods of time. We also presented a pipeline to process data from individual modalities for the applications of classifying hazardous vs. non-hazardous incidents while driving. Furthermore, we presented a method to be able to use pre-trained convolution neural networks to extract deep-learning based features from these modalities in addition to traditionally used ones. In future, we would explore the addition of eye-tracking as another sensor modality to study gaze dynamics and automatic stimuli tagging.

\section*{ACKNOWLEDGMENT}
We would like to thank our colleagues from UCSD LISA Lab who helped us with data collection. We would also like to thank the research groups associated with collecting and disseminating the KITTI Dataset.

\end{document}